\documentclass[11pt,twoside]{article}
\usepackage{asp2010}
\usepackage{natbib}
\usepackage{url}
\def\jgr{J.~Geophys.~Res.~}
\def\apj{Astrophys.~J.~}
\def\apjs{Astrophys.~J.~Suppl.~S.~}

\resetcounters

\begin{document}

\title{Latitudinal Connectivity of Ground Level Enhancement Events}
\author{N.~Gopalswamy$^{1}$, P.~M\"akel\"a$^{1,2}$ \affil{$^1$NASA Goddard Space Flight Center, Greenbelt, MD
20771, U.S.A.\\ $^2$The Catholic University of America, Washington, DC
20064, U.S.A.}}

\begin{abstract}
We examined the source regions and coronal environment of the historical ground level enhancement (GLE) events in search of evidence for non-radial motion of the associated coronal mass ejection (CME). For the 13 GLE events that had source latitudes $>$30$^\circ$ we found evidence for possible non-radial CME motion due to deflection by large-scale magnetic structures in nearby coronal holes, streamers, or pseudo streamers. Polar coronal holes are the main source of deflection in the rise and declining phases of solar cycles. In the maximum phase, deflection by large-scale streamers or pseudo streamers overlying high-latitude filaments seems to be important. The B0 angle reduced the ecliptic distance of some GLE source regions and increased in others with the net result that the average latitude of GLE events did not change significantly.  The non-radial CME motion is the dominant factor that reduces the ecliptic distance of GLE source regions, thereby improving the latitudinal connectivity to Earth. We further infer that the GLE particles must be accelerated at the nose part of the CME-driven shocks, where the shock is likely to be quasi-parallel.
\end{abstract}

\section{Introduction}
Ground level enhancement (GLE) in solar energetic particle (SEP) events represents the highest energy ($\sim$GeV) particles accelerated during large solar eruptions.  GLE particles are thought to be accelerated by the flare process or the CME-driven shock.  Extensive CME observations became available only during Solar Cycle (SC) 23 (Cliver 2006; Gopalswamy et al.\ 2012a). Among the 16 GLEs observed during SC 23, fast and wide CMEs were observed in all but one GLE\@. The 1998 August 24 GLE was the exception,  occurring when the SOHO spacecraft was temporarily disabled (no CME observations). However, a fast interplanetary CME was observed, implying a fast CME near the Sun. Thus a one-to-one correspondence between GLEs and fast CMEs was established (Gopalswamy et al.\ 2012a). Based on metric type II radio burst onset, flare rise time, and CME speed, these authors found that the CME shocks form very close to the Sun ($<$0.5 solar radii above the solar surface) and that the shocks have sufficient time to accelerate particles to GeV energies. These observations strongly support the shock acceleration mechanism. Even though a large number of type II radio bursts were observed during SC 24 (indicating the occurrence of CME-driven shocks), there was only one GLE (on May 17, 2012) so far. Gopalswamy et al. (2013) suggested that poor latitudinal connectivity may be one of the main reasons for the very low occurrence rate of GLEs. In this paper, we present additional evidence in support of this connectivity constraint by considering all GLE events since their discovery in 1942 (Forbush, 1946). Heliographic coordinates of the associated flares are known for almost all CMEs (Cliver et al.\ 1982; Cliver, 2006; Gopalswamy et al.\ 2012a; Nitta et al.\ 2012; Miroshnichenko et al.\ 2013). CMEs were discovered in 1971, but routine information became available only after the launch of SOHO in December 1995.  Therefore, we cannot determine CME non-radial motion for most GLEs. However, we identify coronal holes and streamers that can deflect the GLE-causing CMEs.
\begin{figure}[!ht]
\plotone{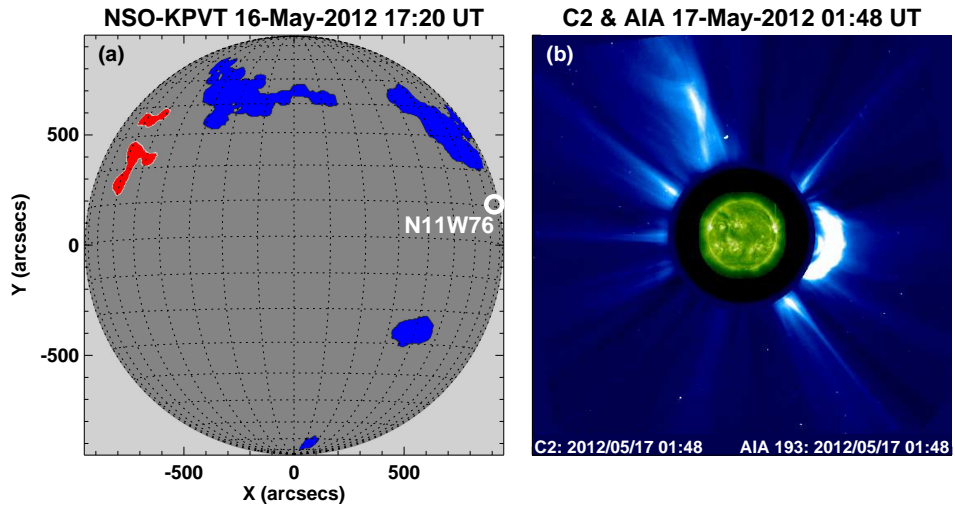}
\caption{(a) A coronal hole map derived from the Kitt Peak Vacuum Telescope (KPVT) data. The blue and red patches correspond to coronal holes with positive and negative polarities, respectively. The flare location (N11W76) of the 2012 May 17 GLE at 02:40 UT is shown by the white circle.  The large coronal hole to the north of the GLE source region is consistent with the southward (non-radial) motion of the GLE-associated CME\@. (b) The white-light CME in a SOHO/LASCO image taken by its C2 telescope at 01:48 UT\@. Superposed on the coronagraph image is a 193 \AA~EUV image taken by the Solar Dynamics Observatory's Atmospheric Imaging Assembly (AIA) showing the flare location close to the northeast limb.}
\end{figure}

\section{The 2012 May 17 GLE Event}
The first and only GLE event of SC 24 as of this writing originated from NOAA active region 11476 in the northwest quadrant of the Sun (N11W76). Coronagraph images from SOHO and STEREO show that the CME moved non-radially, heading in the southwest direction (Gopalswamy et al.\ 2013).  Forward-fitting of the coronagraph images using Thernisien (2011) flux rope model indicated that the effective location of the CME source region was S07W76, indicating that the CME was deflected to the south by about 18$^\circ$. When we examined the coronal environment of the GLE source a few hours before the eruption, we found a large coronal hole close to the active region in the northeast direction (see Fig.~1a).  The southern end of the coronal hole was less than 10$^\circ$ away from the active region. The approximate centroid of the coronal hole was at N30W60. The CME motion is thus consistent with a deflection by the coronal hole (Fig.~1b).  Such deflections are thought to cause ``driverless'' shocks observed at Earth in extreme cases (Gopalswamy et al.\ 2009; 2010) and decide whether a CME flux rope appears as a magnetic cloud or non-cloud ejecta in in-situ observations (Xie et al.\ 2013; M\"akel\"a et al.\ 2013). The configuration in Fig.~1 is the motivation for us to look at historical GLE events for possible non-radial motion of the associated CMEs.
The SC 24 GLE was associated with only an M5.1 flare, which is well below the typical soft X-ray flare size for SC 23 GLEs (X3.8). Historically, there were only two other GLEs with flare size smaller than the SC 24 GLE: GLE 33 on 1979 August 21 with a C6 flare and GLE 35 on 1981 May 10 with a M1 flare (Cliver, 2006; Gopalswamy et al.\ 2012a). The CME associated with the SC 24 GLE was very fast (~2000 km/s), capable of driving a strong shock, and hence produce GLE particles (Gopalswamy et al.\ 2013).  By examining the source regions of several eruptions from SC 24 that had the same the source longitude range as this GLE event, Gopalswamy et al.\ (2013) concluded that the non-GLE source regions had latitudinal distance from the ecliptic: 32$^\circ$ compared to 9$^\circ$ for SC 23 GLEs from similar source longitude (13$^\circ$ when all SC 23 GLEs were considered).  Unfavorable solar B0-angles and non-radial CME motions were found to be the reasons for the increased distance from the ecliptic. The B0 angle is the heliographic latitude of the central point of the solar disk and can vary from -7$^\circ$.23 to + 7$^\circ$.23. B0 represents the fact that the ecliptic plane (where Earth is located) and the Sun's equatorial plane are not aligned. Thus for a northern (southern) CME source with negative (positive) B0 angle the magnetic connectivity to Earth worsens. B0 was -2$^\circ$.4 for the SC 24 GLE, making the latitudinal distance to the ecliptic as 4$^\circ$.6, which is less than the average distance for SC 23 GLE source regions (13$^\circ$). Thus the latitudinal connectivity of the shock nose is important for GLEs.

\section{Source Regions of Historical GLE Events}
The historical GLE events are shown as a ``butterfly'' diagram in Fig.~2. The source locations were taken from the published literature (see e.g., Cliver et al.\ 1982; Cliver 2006; Gopalswamy et al.\ 2012a; Miroshnichenko et al.\ 2013).  The first two GLEs were observed in SC 17 on 1942 February 28 and March 7, respectively. Forbush (1946) reported on these two GLEs and on the third GLE (1946 July 25), which occurred in SC 18. For the five cycles starting from SC 19, there were roughly a dozen GLEs per cycle, but we have only one GLE event in SC 24.  Figure~2b shows that the source latitudes are below 40$^\circ$, as expected because the energy needed for the GLE events are available only in sunspot regions. The latitude distribution is bimodal because of the northern and southern active region belts. The average latitudes are -15$^\circ$.3 (south) and 19$^\circ$.6 (north). When we corrected for the B0 angle, the average ecliptic distances did not change significantly: -17$^\circ$.9 and 19$^\circ$.2 in the southern and northern hemispheres, respectively. These average latitudes are slightly larger than the 13$^\circ$ obtained for SC 23 events (after accounting for non-radial CME motion). Unfortunately, we have no CME observations for most of the pre-SOHO GLEs, but we shall show that the effective ecliptic distances are expected to be lower for pre-SOHO GLEs based on their coronal environment. The source longitudes of historical GLE events (Fig.~2d) range from E88 to W150, but there were only two events beyond E15 (E88 and E39). On the other hand, there were a dozen events behind the west limb. The longitude distribution suggests that the magnetic connectivity of Earth to the source region is very important.
\begin{figure}[!ht]
\plotone{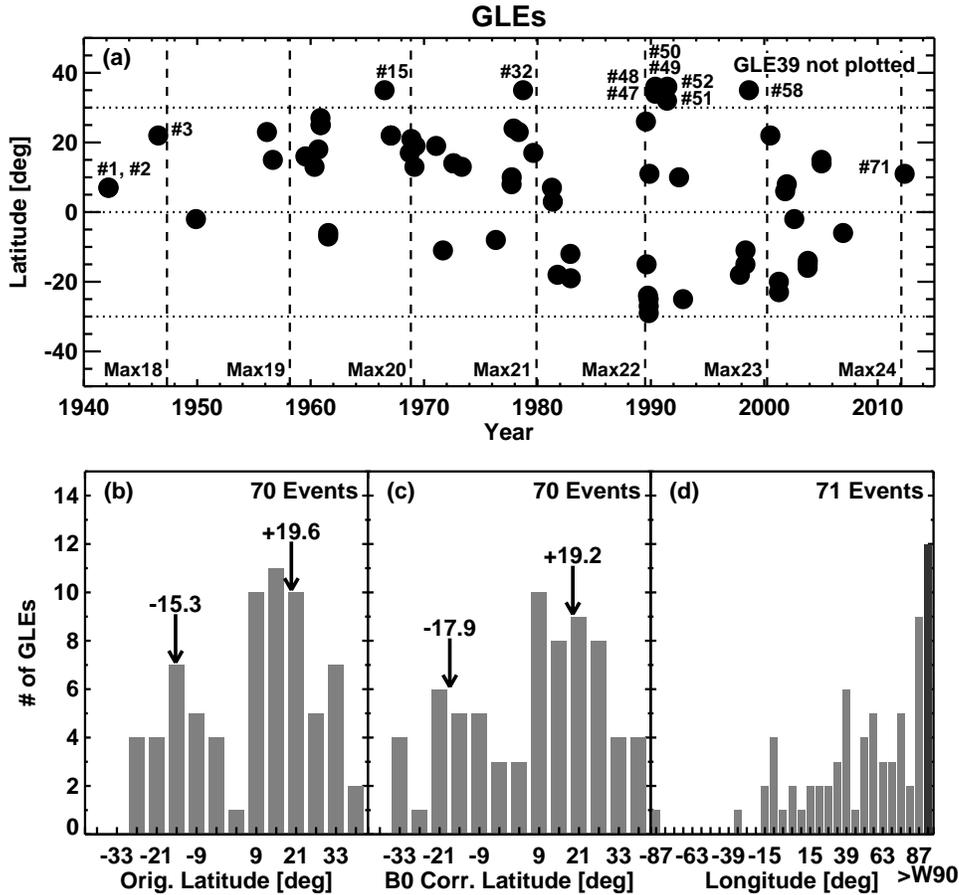}
\caption{(a) GLE latitude as a function of time for GLEs since their discovery in 1942. The approximate times of solar maxima of cycles 18--24 are denoted by vertical dashed lines (from http://sidc.oma.be/DATA/monthssn.dat). (b) The distribution of flare latitudes of the GLE events as in (a). (c) The  latitude distribution corrected for the B0 angle.  (d) The distribution of flare longitudes. The bin size is 6 degrees for all the distributions. In the latitude distribution, there are only 70 events because the latitude of the backside event 39 is unknown. The two GLEs with the easternmost source longitudes are on 1960 September 3 and 1978 April 29.  In the longitude distribution, the $>$90$^\circ$ bin (the dark bin) includes all behind the west-limb events.
}
\end{figure}

\subsection{Higher-Latitude GLEs}
The GLE events with latitudes $>$30$^\circ$ are of interest because they clearly contradict our conclusion that the ecliptic distance of the source regions needs to be small for a GLE event. These higher-latitude GLEs are listed in Table 1 along with the flare location, B0 angle, latitudinal distance to the ecliptic ($\lambda$e), Carrington Rotation (CR) number with its starting day, and whether a deflecting structure (coronal hole, streamer, or pseudo streamer) existed poleward of the source region. We have included GLEs with source latitude $>$30$^\circ$ either before accounting for B0 angle or after. Before B0 correction, there were nine events with source latitude $>$30$^\circ$ and all were from the north. After B0 correction, GLEs 32 and 58 had $\lambda$e $<$30$^\circ$. GLE 15 had similar trend, but $\lambda$e $\sim$31$^\circ$. After B0 correction, 4 events from the south (GLEs 42--45) had $\lambda$e $>$30$^\circ$ (i.e, the connectivity worsened). The B0 angle was less than 2$^\circ$ and hence insignificant for six events (GLEs 47--52).

GLEs with either flare latitude or $\lambda$e exceeding 30$^\circ$ fall into three groups depending on the SC phase: GLEs 15, 32, and 58 were rise-phase events, GLEs 51 and 52 were declining-phase events, and the remaining 8 were maximum-phase events. Accordingly, one expects different coronal environment for the two groups. For example, strong polar coronal holes (PCH) prevail in the rise phase and the sunspots appear at relatively higher latitudes.

\begin{table}[!ht]
\caption{Historical GLE events with flare latitudes $>$30$^\circ$}
\smallskip
\begin{center}
{\small
\begin{tabular}{clllllll}
\tableline
\noalign{\smallskip}
GLE \# & Date & Flare Loc. & B0 & $\lambda$e & CR \# & Start Date & Deflector\\
\noalign{\smallskip}
\tableline
\noalign{\smallskip}
15 & 1966/07/07	& N35W48	& +3.55	& N31	& 1509 & 22.0747 & PCH?\\
32 & 1978/09/23	& N35W50	& +7.02	& N28	& 1673 & 20.0572 & PCH\\
42 & 1989/09/29	& S24W105	& +6.80	& S31	& 1820 & 11.5184 & S+CH\\
43 & 1989/10/19	& S25E09	& +5.54	& S31	& 1821 & 8.7953	& S\\
44 & 1989/10/22	& S27W32	& +5.29	& S32	& 1821 & 8.7953	& S\\
45 & 1989/10/24	& S29W57	& +5.11	& S34	& 1821 & 8.7953	& S\\
47 & 1990/05/21	& N34W37	& -1.94	& N36	& 1829 & 15.2476 & PS\\
48 & 1990/05/24	& N36W76	& -1.59	& N38 	& 1829 & 15.2476 & PS\\
49 & 1990/05/26	& N35W103	& -1.35	& N36	& 1829 & 15.2476 & PS \\
50 & 1990/05/28	& N35W120	& -1.11	& N36	& 1829 & 15.2476 & PS\\
51 & 1991/06/11	& N32W15	& +0.57	& N31	& 1843 & 1.0619	& PCH\\
52 & 1991/06/15	& N36W70	& +1.07	& N35	& 1843 & 1.0619	& PCH\\
58 & 1998/08/24	& N35E09	& +7.01	& N28 	& 1939 & 1.3214	& PCH\\
\noalign{\smallskip}
\tableline
\end{tabular}
}
\end{center}
\end{table}

We examined coronal hole maps available from Kitt Peak National Observatory (KPNO): \url{ftp://nsokp.nso.edu/kpvt/synoptic/choles/} for Carrington Rotations 1633 (1975 September 25) to 1987 (2002 March 2).  We also used Yohkoh soft X-ray data for confirmation in SC 23. Finally, we also examined the H-alpha synoptic charts (\url{ftp://ftp.ngdc.noaa.gov/STP/SOLAR_DATA/SGD_PDFversion/})from the Solar Geophysical Data to understand the environment of GLE source regions. Figure~3 shows the source regions of GLEs 51, 52, and 58 on He 10830 \AA~coronal hole maps. Clearly there were polar coronal holes (PCH) in each of the three cases. For GLE 58, there is additional confirmation from Yohkoh Soft X-ray Telescope (SXT) images, which show a prominent PCH in the north (see also Fig.~3 in Gopalswamy et al.\ 2012a). KPNO maps show north PCH for GLE 32 also. Sheeley (1980) reported on this PCH (his Figure~1), which shows an extension of the PCH in the direction of the GLE source.  Note that GLEs 51 and 52 occurred right after the polarity reversal in the northern hemisphere during cycle 22 and the PCH started increasing area. On the other hand GLE 58 occurred during the rise phase of SC 23, when the PCH area is close to its peak.

\begin{figure}[!ht]
\plotone{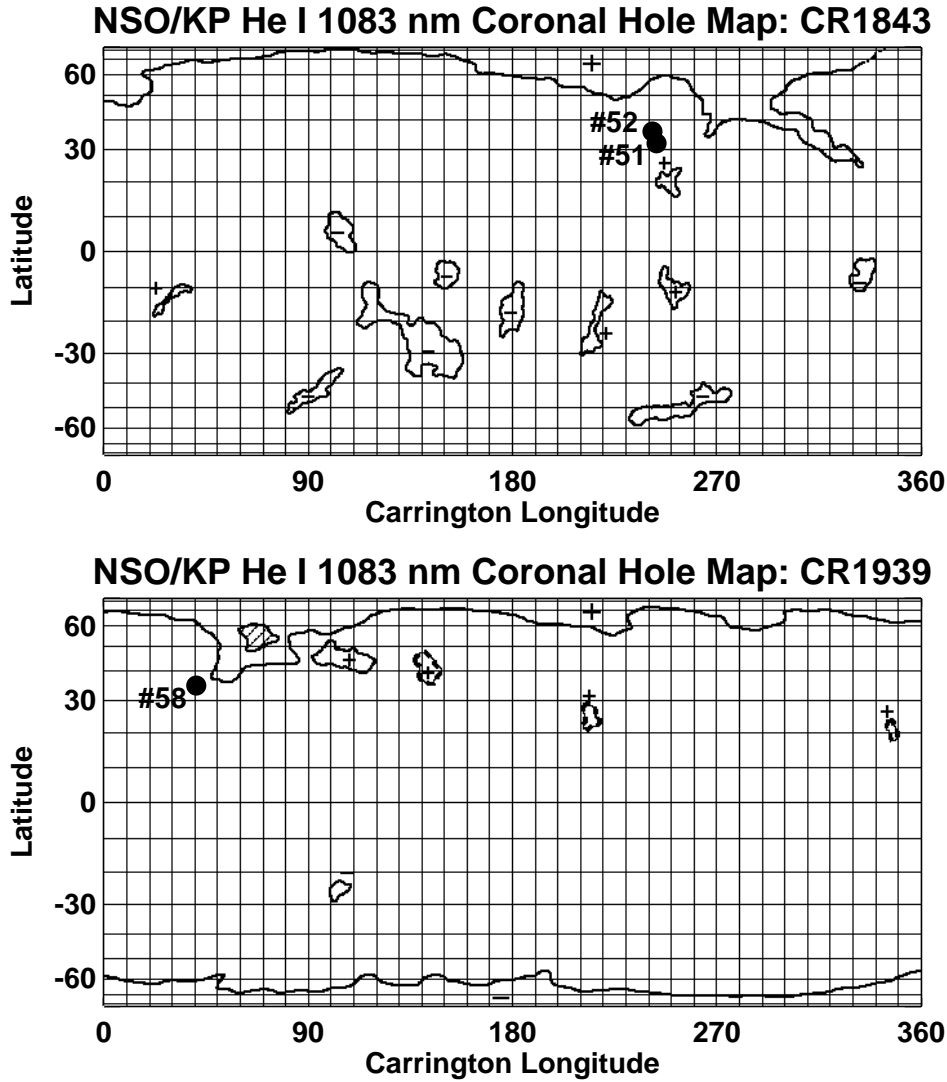}
\caption{Coronal hole synoptic maps with the GLE source regions superposed for Carrington rotations 1843 (GLEs 51, 52) and 1939 (GLE 58). The maps were derived from He 1083 nm spectroheliograms at KPNO.}
\end{figure}

There was no coronal hole observations for GLE 15, but one can infer a PCH because the event occurred during the rise phase of SC 20. The cycle started in October 1964 and the GLE occurred within 2 years into the cycle. The situation is somewhat similar to GLE 58, which also occurred within 2 years into SC 23. The PCH attains its maximum area around the solar minimum, remaining high for two years on either side of the minimum.  The GLE sources are generally at higher-latitude during the rise phase because sunspots originate at higher latitudes in the rise phase. Thus the combination of strong polar coronal holes and the higher-latitude source regions is conducive for CME deflections. This has been suggested as the reason for the offset between position angles of prominence eruptions and the corresponding CMEs (Gopalswamy et al.\ 2003; 2012b) and the larger fraction of flux-rope type CMEs (magnetic clouds) occurring during the rise phase of solar cycles (Gopalswamy et al.\ 2008). Thus we conclude that GLE 15 is consistent with a possible PCH deflection of the associated CME.

The eight maximum-phase events occurred during SC 22. Since the PCHs disappear in this phase, we do not expect deflection of CMEs by coronal holes. Figure~4 shows that there was no coronal hole but an extended ``switchback'' filament to the north of the source region of GLEs. The leading and trailing branches of the switch back filaments can also be seen in the He 10830 \AA~synoptic map in Fig.~4. H-alpha synoptic chart available at the Solar Geophysical Data confirms the switch back filament (see the H-alpha solar synoptic chart for Carrington Rotation 1829 in \url{ftp://ftp.ngdc.noaa.gov/STP/SOLAR_DATA/SGD_PDFversion/}). The filament was very long, extending from Carrington longitude 170$^\circ$ to 360$^\circ$. The trailing and leading branches of the switchback were about 15$^\circ$ and 35$^\circ$ from the active region. One expects a pseudo-streamer type magnetic configuration immediately to the north of the eruption region. We suggest that the magnetic field of the streamer behaves similar to that of coronal holes in deflecting CMEs. For GLEs 42--45, the situation was very similar in the southern hemisphere. GLE 42 occurred from the same source region of GLEs 43--45, but one Carrington rotation earlier. In addition the long filament, GLE 42 had an isolated coronal hole to the southwest, which also might have contributed to the deflection (see H-alpha solar synoptic chart for Carrington Rotation 1820 and 1821 in \url{ftp://ftp.ngdc.noaa.gov/STP/SOLAR_DATA/SGD_PDFversion/}). The switch back was not as sharp as in Fig.~4, but the filament was even longer: 135$^\circ$ to 360$^\circ$ in Carrington longitude. The streamer nearest to the active region and overlying the filament is expected to be a normal streamer. Thus we conclude that the coronal environment is conducive for an equatorward deflection in all the GLEs listed in Table 1. The last column of Table 1 lists the magnetic structures that might have caused the deflection: a polar coronal hole (PCH), streamer (S), or pseudo streamer (PS). In one case, it is possible that a streamer and a coronal hole might have jointly caused the deflection.
A preliminary examination of the remaining GLE events with CH observations indicate that small coronal holes were present at large distances, suggesting little influence on the CMEs. In some cases, the coronal holes were favorably located to improve the connectivity.  There were a few cases in which the deflection would worsen the connectivity, but these need a detailed investigation to see if the ecliptic distance would increase substantially. A detailed report will be published elsewhere.
\begin{figure}[!ht]
\plotone{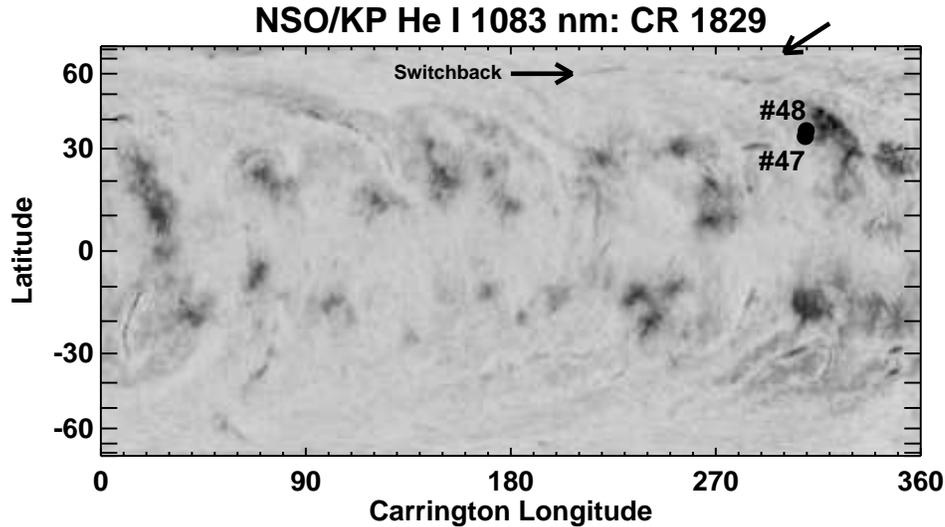}
\caption{He 10830 \AA~synoptic map with the source region of GLEs 47 and 48 marked. In the map compact dark patches are active regions and thin elongated features are filaments. A long switch back filament can be seen to the north of the GLE source region. The arrow with the label "switchback" shows roughly the location where the two arms converge. The poleward arm is pointed by the upper arrow.}
\end{figure}

\section{Discussion}
The main results of this paper are: (i) the non-radial motion of the SC 24 GLE seems to be due to the deflection by a coronal hole located to the northeast of the eruption region, (ii) the CMEs in historical GLE events with flare latitude $>$30$^\circ$ occurring in the rise and declining SC phases seem to be deflected toward the ecliptic by polar coronal holes, (iii) higher-latitude GLEs occurring in the maximum phase seem to deflected by a large-scale streamer or pseudo streamer structure overlying high-latitude filaments. The B0 angle reduced the ecliptic distance of the GLE source region in some cases and increased in others with the net result that the average latitude of GLE events did not change significantly. The non-radial motion seems to the dominant factor in reducing the ecliptic distance of GLE source regions and hence increasing the latitudinal connectivity to Earth.
The coronal deflection of CMEs happens because of the enhanced magnetic content of the coronal holes (Gopalswamy et al.\ 2009; Gopalswamy et al.\ 2010; Shen et al.\ 2011). For polar coronal holes, this represents the solar dipolar field, which is the strongest during solar minima (see e.g., Svalgaard et al.\ 1978; Gopalswamy et al.\ 2003). One of the main properties of coronal holes is that the magnetic field at the photospheric level is enhanced and unipolar relative to the neighboring quiet regions. The field expands into the corona and represents magnetic pressure gradient between the coronal hole and the eruption region (e.g., Panasenco et al.\ 2012; Kay et al.\ 2013) and hence pushes the CME away from the coronal hole. We suggest that the same physical picture applies when a large-scale streamer or pseudo streamer is present on one side of the eruption region causing the magnetic pressure gradient.
The equatorward deflection of CMEs associated with GLE events suggests that Earth may be connected to the nose part of the CME-driven shocks.  Given the fact that GLE particles are released when the CME leading edge is at a heliocentric distance of $\sim$3 Rs (Gopalswamy et al.\ 2012a), we suggest that the GLE particles may be accelerated at the quasi-parallel section of CME-driven shocks. This is because the nose part is well above the source surface where the field lines in the ambient medium are expected to be radial.
It must be pointed out that coronal hole deflection may not be the only reason for non-radial CME motion.  In some active regions, it is possible that non-radial ejections happen due to the magnetic complexities in the source regions. It is also possible that CMEs involve a high-inclination flux rope so that the nose region is much extended in the north-south direction.
We confirm that the ecliptic distance of the source region of a solar eruption is one of the factors that determine whether an SEP event becomes a GLE event. This may be one of the reasons why GLE events are so rare. Other reasons include the reduced number of energetic eruptions during SC 24 and the reduction in seed particles (and preceding CMEs) available to be accelerated by the CME-driven shocks. The reduced efficiency of particle acceleration by the shocks due to the change in physical conditions in the heliosphere (e.g., increase in the Alfv\'en speed of the ambient medium). Thus, GLE events require special conditions in terms of CME kinematics, coronal environment, and magnetic connectivity to Earth.

\section{Summary}
We have confirmed that non-radial CME motion is likely to have happened in historical GLE events that occurred at latitudes $>$30$^\circ$ due to deflection by large-scale magnetic structures in coronal holes or in streamers.  We also infer that the highest energy particles are produced at the nose part of the CME-driven shocks, where the shock strength is highest. Furthermore, the shock is likely to be quasi-parallel in the nose region because the GLE associated CMEs typically cross the potential field source surface at the time of GLE particle release. The special conditions needed to detect a GLE event at Earth coupled with the reduced frequency of energetic eruptions may be responsible for the paucity of GLE events in SC 24.

\acknowledgements This work utilizes SOLIS data obtained by the NSO Integrated Synoptic  Program (NISP), managed by the National Solar Observatory, which is operated by the Association of Universities for Research in Astronomy (AURA), Inc.\ under a cooperative agreement with the National Science Foundation. Work supported by NASA's Living with a Star program.

\end{document}